\documentclass[%
10pt,
twocolumn,
twoside,
superscriptaddress,
nofootinbib,
amsmath,
amssymb,
aps,
prd,
]{revtex4-2}

 \usepackage{multirow}
\usepackage{graphicx}
\usepackage{dcolumn}
\usepackage{bm}

\usepackage{xcolor}

\makeatletter
\def\tagform@#1{(\textcolor{magenta}{#1})}
\makeatother
\usepackage[colorlinks=true,linkcolor=magenta,citecolor=magenta]{hyperref}%

\usepackage{nicefrac}
\usepackage{lipsum}
\usepackage[normalem]{ulem}

\begin{document}

\preprint{FDRXXX}

\title{\color{teal}
Functional Dimensional Regularization for the Lee-Yang Model}

\author{P. Beretta}
\email{pberetta@fing.edu.uy}
\affiliation{IFFI, Facultad de Ingenier\'ia, Universidad de la Rep\'ublica, J.H.y Reissig 565, 11000 Montevideo, Uruguay}
\affiliation{IFFC, Facultad de Ciencias, Universidad de la Rep\'ublica, Iguá 4225, 11400 Montevideo, Uruguay}

\author{C. A. S\'anchez-Villalobos}
\email{carlos.sanchez@fcien.edu.uy}
\affiliation{IFFC, Facultad de Ciencias, Universidad de la Rep\'ublica, Iguá 4225, 11400 Montevideo, Uruguay}

\author{A. Codello}
\email{alessandro.codello@unive.it}
\affiliation{DSMN, Ca' Foscari University of Venice, Via Torino 155, 30172 - Venice, Italy}
\affiliation{IFFI, Facultad de Ingenier\'ia, Universidad de la Rep\'ublica, J.H.y Reissig 565, 11000 Montevideo, Uruguay}

\date{\today}

\begin{abstract}
We apply Functional Dimensional Regularization (FDR) to the \texttt{Lee-Yang} universality class, a paradigmatic non-unitary theory, and compute its main critical exponents across the dimension range $1\leq d\leq 6$. By combining a second-order derivative expansion with polynomial field expansions, we obtain results that are compatible with resummed $\varepsilon$-expansion as well as with state-of-the-art estimates from Conformal Bootstrap, Fuzzy Sphere and Non-Perturbative Renormalization Group. This demonstrates that, in the context of non-unitary theories, FDR retains the efficiency and rapid convergence to accurate critical exponents already observed in the \texttt{Ising} and \texttt{O(N)} cases. 
\end{abstract}

\maketitle


\subsection*{\color{teal}Introduction}\vspace{-0.2cm}

The \texttt{Lee-Yang} universality class is a cornerstone of non-unitary critical phenomena \cite{Cardy:2023lha}, and it is closely connected to both the Lee-Yang zeroes and the Yang-Lee singularities \cite{yang1952statisticalI,yang1952statisticalII,Fisher:1978pf,cardy2024yang}. The Lee-Yang zeroes are the roots of the partition function, and studying them provides a useful way to analyze the phase diagram in the thermodynamic limit, whereas the Yang-Lee singularities are points in the equation of state of the order parameter that behave as critical points with associated universal features. The \texttt{Lee-Yang} class appears when studying the Yang-Lee singularity in the Ising model, and Fisher established that it can be described by a scalar field theory with a cubic interaction, which indicates an upper critical dimension of six \cite{Fisher:1978pf}. This corresponds to one of the few non-trivial models that remain interacting above four dimensions, which makes it a crucial testing ground for theoretical methods. Furthermore, in the context of conformal field theory (CFT), Cardy identified it with the non-unitary minimal model $\mathcal{M}_{2,5}$ in two dimensions \cite{Cardy:1985yy}.

The relevance of the \texttt{Lee-Yang} properties extends to a variety of physical systems (see \cite{cardy2024yang} for a recent review). For repulsive lattice gases and continuous gases there is a negative-pressure singularity with critical data in the \texttt{Lee-Yang} class \cite{cardy1982directed,dhar1983exact,poland1984universality,baram1987universality}. Parisi and Sourlas proposed that the $d$-dimensional \texttt{Lee-Yang} universal properties are related to the $d+2$ scaling exponent of branched polymers \cite{parisi1981critical}, which was analytically supported in \cite{brydges2003branched}. The universal location of the Yang-Lee singularity has been studied to determine the radius of convergence of the small chemical potential expansion in QCD \cite{stephanov2006qcd,Rennecke:2022ohx}.

Despite its relevance, computing precise \texttt{Lee-Yang} critical data in spatial dimensions $d=3,4,5$ remains a challenge. This has been tackled by several methods, which require involved computations. The $\varepsilon = 6 -d$ expansion ($\varepsilon$E), computed to six loops by Schnetz \cite{Schnetz:2025wtu}, requires sophisticated Borel-Pad\'e resummation techniques \cite{Borinsky:2021jdb,Gracey:2025rnz}. Conformal Bootstrap  (CB) methods, while powerful for unitary theories, are less effective here \cite{Gliozzi:2013ysa,Gliozzi:2014jsa,Hikami:2017hwv}. The activity expansion of the hard dimer model (HT), related to the high-temperature expansion of the Ising model, has also been employed up to its 24th order \cite{Butera:2012tq}.
Non-Perturbative Renormalization Group (NPRG) methods have been used \cite{li2011functional,An:2016lni,Zambelli:2016cbw,Kockler:2026euo}, and although obtaining accurate critical exponents in three dimensions has proven computationally intensive, these methods have been very successful in determining the edge singularity \cite{Connelly:2020gwa,Rennecke:2022ohx,Johnson:2022cqv}.
In the past few years, the Fuzzy Sphere (FS) technique has emerged as a promising non-perturbative approach in three dimensions \cite{Fan:2025bhc,ArguelloCruz:2025zuq,EliasMiro:2025msj}.

In recent work, a novel functional RG approach has been proposed for the study of critical phenomena: Functional Dimensional Regularization (FDR) \cite{Beretta:2026zcy}. This approach is a functional reformulation of dimensional regularization that goes beyond the classical $\varepsilon$-expansion by incorporating information from all upper critical dimensions. As a result, it yields versatile RG flow equations that are applicable in arbitrary dimension. Combining FDR with the derivative expansion at second order (DE2), the universal critical exponents of the \texttt{Ising} \cite{FDR} and \texttt{O(N)} \cite{Beretta:2026avu} models were computed. The resulting exponents are compatible with those computed using NPRG to the next order of the derivative expansion~(DE4). These successes have inspired us to undertake a testing campaign for FDR in several models that have already been studied with other methods. The goal is to establish the capabilities and limitations of FDR, with the hope that it can be used to tackle problems where other methods have failed. This article is part of that campaign.

In this paper, we apply FDR to study the \texttt{Lee-Yang} universality class and to compute its critical exponents. Using DE2 and a polynomial truncation in the field, we show that FDR yields critical exponents in dimensions $d=5,4,3$ that are competitive with those from state-of-the-art methods. In addition, we show that FDR reproduces the one-loop $\varepsilon$-expansion results and study the full RG spectrum continuously with the dimension in the range $1 \leq d \leq 6$. The extension below three dimensions allows a direct comparison with the exact results in $d=2$ and $d=1$, which provides a stringent test of the method.


\subsection*{\color{teal}Flow equations }\vspace{-0.2cm}


The \texttt{Lee-Yang} universal behavior can be described by a Ginzburg-Landau (GL) theory that is symmetric under $\mathcal{PT}$ transformations \cite{Fisher:1978pf,bender2004extension}.
To study the universal exponents predicted by this GL theory, we employ the FDR scheme together with the derivative expansion up to DE2. The most general DE2 ansatz for the $\mathcal{PT}$-symmetric model is the same as that for a scalar field theory with $\mathbb{Z}_2$ symmetry, except that the potential and the wave-function renormalization are now complex functions. The corresponding flow equations were determined in \cite{FDR}; they express the flow of the dimensionless potential $v(\phi)$ and of the dimensionless wave-function renormalization $z(\phi)$ in closed analytic form, where both $v$ and $z$ are complex functions of the dimensionless field $\phi$. After an appropriate rescaling that eliminates factors of $4\pi$, and using the standard notation $\beta_f = \mu \partial_\mu f$ with running momentum scale $\mu$, the renormalization group flow equations are:
\begin{widetext}
\begin{eqnarray}
\beta_v &=& -d\, v+\tfrac{d-2+\eta}{2}\phi \,v'+e^{-\frac{v''}{z}} \nonumber\\
\beta_z &=& \eta\, z+\tfrac{d-2+\eta}{2}\phi \,z'+ \Big\{-\frac{(v''')^2}{6 z^2}-\frac{z''}{z}
+\left[\frac{1}{z^2}+\frac{v''}{3z^3}\right] v''' z' 
+\left[\frac{3}{2z^2}-\frac{v''}{z^3} -\frac{(v'')^2}{6z^4}\right] (z')^2 \Big\}\, e^{-\frac{v''}{z}} \,.\label{betavz}
\end{eqnarray}
\end{widetext}
%
Note that these equations are valid in any dimension, with their $d$-dependence appearing only in the classical scaling part.

Following what has been done in \cite{Zambelli:2016cbw,An:2016lni,Benedetti:2026tpa,Rennecke:2022ohx} in the context of the NPRG, we make an expansion around a value of the field $\phi_0$ of the form:
\begin{eqnarray}
v(\phi)&=&\tfrac{1}{2!}\lambda_{2}(\phi-i \phi_0)^2+\tfrac{1}{3!}i\lambda_{3}(\phi-i \phi_0)^3\label{expansionv}\\&&+\tfrac{1}{4!}\lambda_{4}(\phi-i \phi_0){}^{4}+\tfrac{1}{5!}i\lambda_{5}(\phi-i \phi_0){}^{5}
+... \nonumber
\\&&\nonumber\\
z(\phi)&=&1+iz_{1}(\phi-i \phi_0)+\tfrac{1}{2!}z_{2}(\phi-i \phi_0){}^{2}\nonumber\\&&+\tfrac{1}{3!}iz_{3}(\phi-i \phi_0)^{3}+... \label{expansionz}\,.
\end{eqnarray}
Here we choose to write all imaginary units explicitly so as to work exclusively with real numbers, since $\mathcal{PT}$-symmetry, acting as $v(\phi)\to v^*(-\phi)$ and $z(\phi)\to z^*(-\phi)$, implies that $\phi_0$, $\lambda_i$, and $z_i$ must be real. We truncate  up to order $N_v$ in \eqref{expansionv} and up to order $N_v-2$ in \eqref{expansionz}, and we increase $N_v$ until we observe convergence to a given accuracy in the critical exponent $\eta$. The running anomalous dimension is defined as usual by imposing the condition $z(i\phi_0)=1$ along the flow; see Eq.~\eqref{etadef} below.

The system of beta functions for $\lambda_i$ and $z_i$ is obtained by inserting \eqref{expansionv} and \eqref{expansionz} into \eqref{betavz} and matching the coefficients of equal powers of the field on the left- and right-hand sides. We employ two types of field expansion that have been used previously in the NPRG context \cite{Zambelli:2016cbw,An:2016lni,Rennecke:2022ohx,Benedetti:2026tpa}: one with $\phi_0 = 0$ (zero-field), and the other around the zero-mass point $v''(i\phi_0) = 0$ (zero-mass). In the latter case, the flow of $\phi_0$ follows from
\begin{align}
\mu\partial_\mu\phi_0 =  i\frac{\left.\mu\partial_\mu v''(\phi)\right|_{\phi = i\phi_0}}{ \left.v'''(\phi)\right|_{\phi = i \phi_0}}\,,
\end{align}
via the condition $\mu \partial_\mu v''(i \phi_0) \equiv 0$.  Although the expansion around $\phi_0\neq0$ is more involved, it has been observed to converge faster than the one with $\phi_0 = 0$ \cite{Zambelli:2016cbw,An:2016lni}. We keep both expansions to check the consistency of these approximations.
Since in the zero-mass expansion the normalization point $i\phi_0$ flows, the condition $z(i\phi_0)=1$ is fixed along the flow,
\begin{equation}
\mu\partial_\mu\, z(i\phi_0) = \beta_z(i\phi_0) + i\, z'(i\phi_0)\, \mu\partial_\mu\phi_0 = 0\,,
\label{etadef}
\end{equation}
which determines $\eta$ as a function of the couplings. The second term in \eqref{etadef} vanishes at the fixed-point (FP), but it contributes to the stability matrix.

Note that the field expansions in \eqref{expansionv} and \eqref{expansionz} are not performed around the minimum of the potential, $v'(\phi_0)=0$. Expanding around this minimum is a method that has produced converged results for the $\mathbb{Z}_2$-symmetric theory in the NPRG \cite{canet2003optimization}; in that case, it was found to converge faster than the expansion with $\phi_0=0$. However, this choice is not appropriate for studying the \texttt{Lee-Yang} class. As noted in \cite{Johnson:2022cqv}, this approach is numerically demanding in an NPRG implementation, because fine-tuning to the Yang-Lee singularity requires a massless theory. We observe the same difficulty with the FDR when solving the flow equations \eqref{betavz}.

To show explicit expressions, we report below the beta functions for the first few couplings in the zero-field expansion. For couplings of the potential we find 
\begin{eqnarray}
\beta_{\lambda_2}	&=&(-2+\eta)\lambda_{2}+\big(-\lambda_{3}^{2}-\lambda_{4}-\lambda_{2}^{2}z_{1}^{2}\nonumber\\&&
+2\lambda_{2}z_{1}^{2}-2\lambda_{3}z_{1}+2\lambda_{2}\lambda_{3}z_{1}+\lambda_{2}z_{2}\big)e^{-\lambda_{2}}\nonumber\\
\beta_{\lambda_3}	&=&\left(\tfrac{d}{2}-3+\tfrac{3}{2}\eta\right)\lambda_{3}+\big(\lambda_{3}^{3}+3\lambda_{3}\lambda_{4}-\lambda_{5}-\lambda_{2}^{3}z_{1}^{3}\nonumber\\&&
+6\lambda_{2}^{2}z_{1}^{3}-6\lambda_{2}z_{1}^{3}+6\lambda_{3}z_{1}^{2}+3\lambda_{2}^{2}\lambda_{3}z_{1}^{2}-12\lambda_{2}\lambda_{3}z_{1}^{2}\nonumber\\&&
+3\lambda_{2}^{2}z_{2}z_{1}+6\lambda_{3}^{2}z_{1}-3\lambda_{2}\lambda_{3}^{2}z_{1}-6\lambda_{2}z_{2}z_{1}+3\lambda_{4}z_{1}\nonumber\\&&
-3\lambda_{2}\lambda_{4}z_{1}+3\lambda_{3}z_{2}-3\lambda_{2}\lambda_{3}z_{2} + z_3 \lambda_2\big)e^{-\lambda_{2}} \nonumber\\
\beta_{\lambda_4}&=& ...\,,\label{betasv}
\end{eqnarray}
while for those of the wave-function renormalization we obtain
\begin{eqnarray}
\beta_{z_{1}} &=&\left(\tfrac{d}{2}-1+\tfrac{3}{2}\eta\right)z_{1}+\big(-\tfrac{1}{6}\lambda_{3}^{3}-\tfrac{1}{3}\lambda_{3}\lambda_{4}+3z_{1}^{3}\nonumber\\&&
+\tfrac{1}{6}\lambda_{2}^{3}z_{1}^{3}+\tfrac{1}{3}\lambda_{2}^{2}z_{1}^{3}-\tfrac{9}{2}\lambda_{2}z_{1}^{3}+\tfrac{9}{2}\lambda_{3}z_{1}^{2}\nonumber\\&&
-\tfrac{1}{2}\lambda_{2}^{2}\lambda_{3}z_{1}^{2}-\tfrac{2}{3}\lambda_{2}\lambda_{3}z_{1}^{2}-\tfrac{1}{3}\lambda_{2}^{2}z_{2}z_{1}\nonumber\\&&
+\tfrac{1}{3}\lambda_{3}^{2}z_{1}+\tfrac{1}{2}\lambda_{2}\lambda_{3}^{2}z_{1}+4z_{2}z_{1}-3\lambda_{2}z_{2}z_{1}\nonumber\\&&
+\lambda_{4}z_{1}+\tfrac{1}{3}\lambda_{2}\lambda_{4}z_{1}+2\lambda_{3}z_{2}+\tfrac{1}{3}\lambda_{2}\lambda_{3}z_{2} -z_3\big)e^{-\lambda_{2}}\nonumber\\
\beta_{z_{2}} &=&...\label{betasz}
\end{eqnarray}
This coupled tower of beta functions can be extended to arbitrarily high order and is equivalent to the full functional system \eqref{betavz} for small field values\footnote{As a check, if we set $\lambda_{2} = m^2$ and $i\lambda_3 = g$ and evaluate equations \eqref{betasv} in $d_c=6$, one recovers the well-known beta functions for these couplings in the scalar $\tfrac{g}{3!}\phi^3$ theory \cite{Borinsky:2021jdb}.}. Finally, the anomalous dimension in the zero-field expansion turns out to be
\begin{equation}
\eta = \Big[
z_{2}
-\tfrac{1}{6}\lambda_{3}^{2}
-\tfrac{\lambda_{2}^{2}+6\lambda_{2}-9}{6}z_{1}^{2}
+\tfrac{\lambda_{2}+3}{3}\lambda_{3}z_{1}
\Big]e^{-\lambda_{2}} \,. \label{eqn:etazero}
\end{equation}
We see that all four $\beta_z$ terms in brackets in Eq.~\eqref{betavz} contribute to the \texttt{Lee-Yang}'s anomalous dimension. 

Once the anomalous dimension is inserted into the flow equations for all coupling constants, imposing the FP condition reduces the problem to an algebraic one.
We locate the \texttt{Lee-Yang} FP of our flow equations iteratively. First, for $N_v = 3$, we can solve the algebraic system analytically in $d = 5.9$. Using this solution as an initial condition, we then lower $d$ in steps $\delta d$ and numerically determine the FP at each step, always using the FP found at dimension $d+\delta d$ as a seed for the one at $d$. For $N_v>3$, we repeat the same procedure, except that the FP at $d = 5.9$ is obtained numerically from the solution with $N_v-1$.
Once a FP solution is found, the RG spectrum follows directly from the stability matrix
\begin{equation}
M_{nm}=\partial \beta_{g_n}/\partial g_m \big|_*\,.
\end{equation}
The eigenvalues $\theta_i$ of this matrix at the FP determine the scaling dimensions of the corresponding universality class.
Within our truncation \eqref{expansionv} and \eqref{expansionz}, the \texttt{Lee-Yang} class corresponds to a FP with a single unstable direction for $d\geq 3$. As we will see below, the stability of the \texttt{Lee-Yang} FP changes for $d<d_*$, where the FP becomes more stable. Ordering the eigenvalues from most relevant to least relevant, $\lbrace \theta_{2}, \theta_{3}, \dots \rbrace$, we have $\theta_{2}<0$ for $d>d_*$, so the correlation-length critical exponent is $\nu = -1/\theta_{2}$, while the correction-to-scaling exponent is $\omega = \theta_{3}$. Below $d_*$ the eigenvalue $\theta_2$ becomes positive and $\nu$ becomes negative, as required by the exact values $\nu^{2d}=-\tfrac52$ and $\nu^{1d}=-1$. In CFT, it is customary to work in terms of the scaling dimensions of the corresponding operators,
\begin{equation}
\Delta_{n} = d + \theta_{n}\,.
\label{eqn:scaling-dimension}
\end{equation}
The \texttt{Lee-Yang} theory is characterized by three related critical exponents: the anomalous dimension $\eta$, the field scaling dimension $\Delta_\phi$, and the exponent $\sigma$ controlling the magnetization divergence at the singularity \cite{Fisher:1978pf}. Depending on the approach, any one of these three quantities can be chosen as the target of the computation, with the other two then obtained from the relations 
\begin{equation}\label{scalingDeltasigma}
\Delta_{\phi} = \tfrac{d-2+\eta}{2}\,,\qquad\qquad \sigma = \tfrac{d-2+\eta}{d+2-\eta} \,.
\end{equation}
In our calculations, the natural choice is $\eta$, as in Eq.~\eqref{eqn:etazero}, but this is not always the case, especially in other approaches.
Exact values for these exponents are known in one ($\eta^{1d}=-1$, $\Delta_\phi^{1d}=-1$, $\sigma^{1d}=-\tfrac{1}{2}$) and two dimensions ($\eta^{2d}=-\tfrac{4}{5}$, $\Delta_\phi^{2d}=-\tfrac{2}{5}$, $\sigma^{2d}=-\tfrac{1}{6}$) \cite{Fisher:1978pf,Cardy:1985yy}, while for $d\geq6$ the mean-field approximation is valid ($\eta^{\rm MF}=0$, $\Delta_\phi^{\rm MF}=\tfrac{d}{2}-1$, and $\sigma^{\rm MF}=\tfrac{1}{2}$).  In this work we cover the whole range $d\in[1,6]$; for $2<d<6$ no exact result is known, apart from the $\varepsilon$-expansion just below six dimensions.
The \texttt{Lee-Yang} correlation-length critical exponent is not independent but is related to the anomalous dimension via the scaling relation
\begin{equation}
\nu = \frac{2}{d-2+\eta} \label{eqn:scaling-relation}\,.
\end{equation}
Finally, the correction to scaling exponent $\omega$ is related to the scaling dimension of the operator $\phi^3$ as in \eqref{eqn:scaling-dimension}.

As previously noted in the NPRG context in \cite{An:2016lni}, at least the second order of the derivative expansion is required to characterize the spectrum of the \texttt{Lee-Yang} class when using a field expansion. The lowest order of the derivative expansion, the LPA, gives the incorrect result $\eta=0$, whereas $\eta$ is already nonzero at one-loop order. Moreover, using FDR we attempted a field expansion at the LPA' level, in which $z$ is independent of the field (in this terminology, the LPA corresponds to setting $z=1$, whereas the LPA' retains a field-independent $z$ but drops the $z$-dependent terms on the right-hand sides of the flow equations for $v$ and $z$, i.e.\ in $\beta_v$ and $\beta_z$). In this case, the results obtained from the expansion around zero-field and from the expansion around zero-mass already differ in $d=5$. The expansion around zero-field yields an $\omega$ that vanishes between $d=5$ and $d=6$, signaling a change in the stability of the FP. The expansion around $v''=0$ yields an $\omega$ that vanishes below $d=4$. However, the resulting values of $\eta$ differ by more than $80\%$ from those obtained with other approaches.
%

\subsection*{\color{teal}Recovering the $\varepsilon$-expansion}\vspace{-0.2cm}

As a first check of our formalism, we recover the standard one-loop results of the $\varepsilon$-expansion from the expansion around zero-field. The system formed by Eqs.~\eqref{betasv} and \eqref{betasz} can be solved analytically near $d=6$. In fact, for the leading-order corrections it is sufficient to consider the LPA' approximation, in which $z_i=0$. Then, following the procedure outlined in \cite{FDR}, which exploits the fact that the beta function of the coupling $\lambda_i$ is linear in $\lambda_{i+2}$, we obtain the following general FP solution in terms of $\eta$ and $\lambda_2$:
\begin{eqnarray}
\lambda_3&=&i\sqrt{6}\sqrt{\eta e^{\lambda_2}}\nonumber\\
\lambda_{4}&=&\eta e^{\lambda_2}(\lambda_2+6)-2e^{\lambda_2}\lambda_2\nonumber\\
\lambda_{5}&=&i\sqrt{\tfrac{3}{2} \eta}e^{\tfrac{3}{2}\lambda_2}(-\varepsilon+27 \eta-6(2-\eta)\lambda_2)\nonumber\\
\lambda_{6}&=&\cdots\,.
\label{eqn:one-loop-epsilon-FP}
\end{eqnarray}
To search for fixed points in the vicinity of the Gaussian, we truncate the solution \eqref{eqn:one-loop-epsilon-FP}, substitute the truncated expression into the remaining beta functions of \eqref{betasv}, and then solve the resulting system under the ansatz $\lambda_2 = \alpha \,\varepsilon$ and $\eta = \gamma\, \varepsilon$. To leading order $O(\varepsilon)$, we then find
\begin{equation}
\lambda_2= -\tfrac{\varepsilon }{3}+O(\varepsilon^2)\,,\quad\quad
\lambda_3 = \sqrt{\tfrac{2}{3}} \sqrt{\varepsilon }+O(\varepsilon^{3/2})\,,
\end{equation}
while all other couplings are of order at least $O(\varepsilon^2)$. We thereby recover the one-loop anomalous dimension $\eta=-\tfrac{\varepsilon}{9}$.
Computing the stability matrix and evaluating at the FP, we recover the \texttt{Lee-Yang} RG spectrum at order $\varepsilon$ \cite{Codello:2017hhh}:
\begin{equation}
\Delta_{\phi^n} = d-6+2 n+\tfrac{1}{9} (3 n^2-8 n+9) \,\varepsilon -\tfrac{\varepsilon}{3}  \delta_{n,3} + O(\varepsilon^2)\,.
\end{equation}
In particular, we recover the well known expressions
$
\nu =\tfrac{1}{2}+\tfrac{5}{36}\varepsilon + O(\varepsilon^2)
$
and 
$ 
\omega = \varepsilon + O(\varepsilon^2)
$\,.
This reasoning cannot be followed with the zero-mass expansion, as in this case $\lambda_2$ is fixed to 0. Nevertheless, as we will see below,  the numerical RG spectrum is aligned with the leading order $\varepsilon$E prediction presented above. 
Thus the FDR flow \eqref{betavz} is correct to leading order in $\varepsilon$ for the \texttt{Lee-Yang} model. 

The Lee-Yang model is not the only non-unitary theory that is nontrivial in three dimensions. The other one is the tricritical Lee-Yang model (\texttt{Tri-Lee-Yang}), which has upper critical dimension $d_c=\tfrac{10}{3}$; all other non-unitary models (with GL Lagrangian $i\phi^{2n+1}$) are trivial in three dimensions \cite{Zambelli:2016cbw,Codello:2017qek}, since their upper critical dimension is smaller than three. The \texttt{Tri-Lee-Yang} universality class has been analyzed perturbatively in $d=\tfrac{10}{3}-\varepsilon$ dimensions at three-loop order \cite{Codello:2017epp,Gracey:2017okb}. It has also attracted attention in the FS community, where it has been connected to the two-dimensional non-unitary minimal model $\mathcal{M}_{2,7}$ \cite{Katsevich:2025ojk}.
We leave the FDR study of the \texttt{Tri-Lee-Yang} class, and of all other non-unitary multicritical models described in Refs.~\cite{Codello:2017epp,Benedetti:2026tpa}, for future work.

\begin{figure}[!h]
\centering
\includegraphics[width=1\columnwidth]{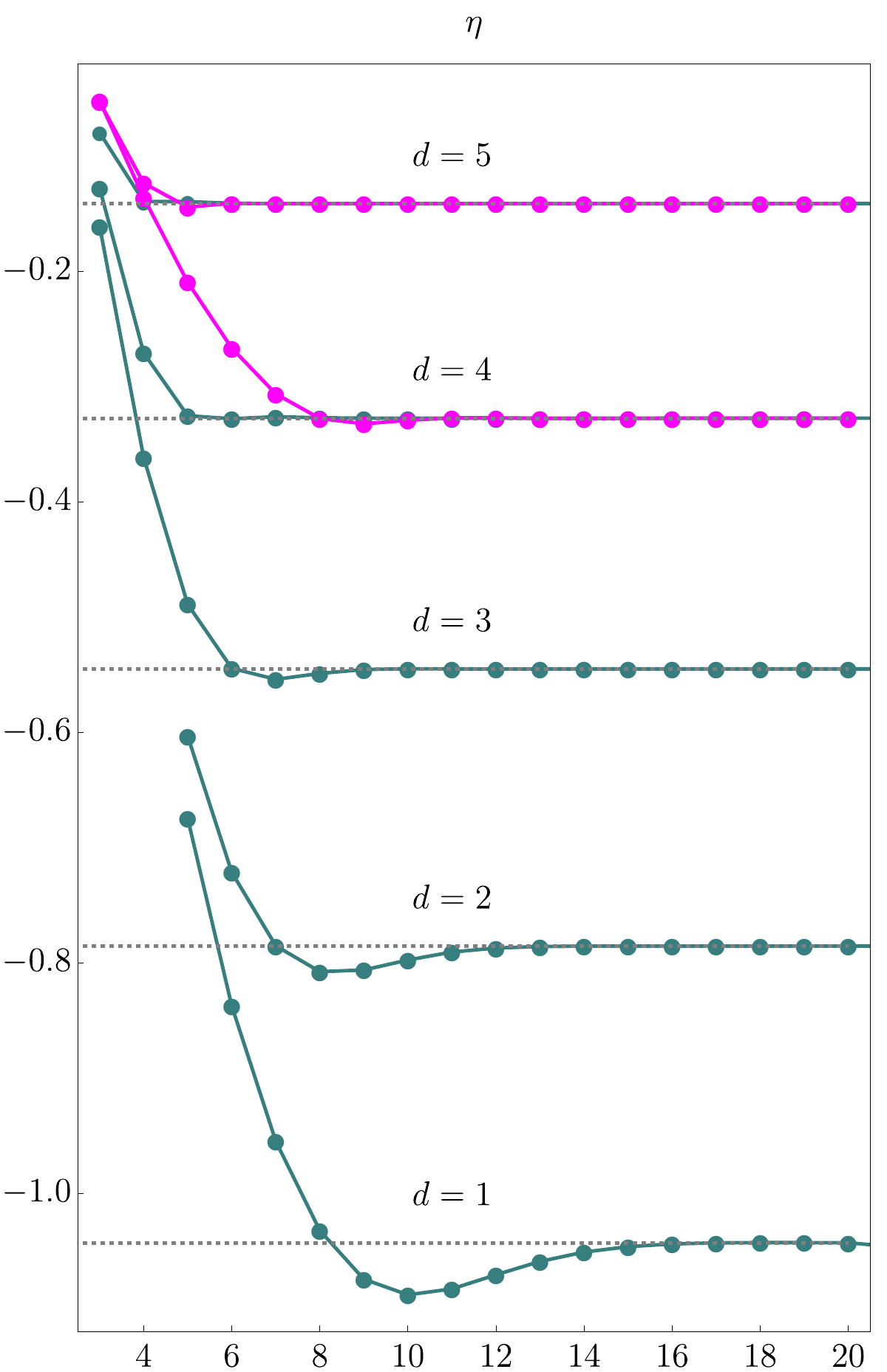}
\caption{Convergence of the anomalous dimension $\eta$ in $d=5,4,3,2$ and $1$ (from top to bottom) as a function of the truncation order $N_v$, comparing the zero-field expansion (magenta) with the zero-mass expansion (teal). Dotted lines mark the converged zero-mass values.}
\label{EtaConvergence}
\end{figure}

\subsection*{\color{teal}Full analysis}\vspace{-0.2cm}

\begin{figure}
\centering
\includegraphics[width=1\columnwidth]{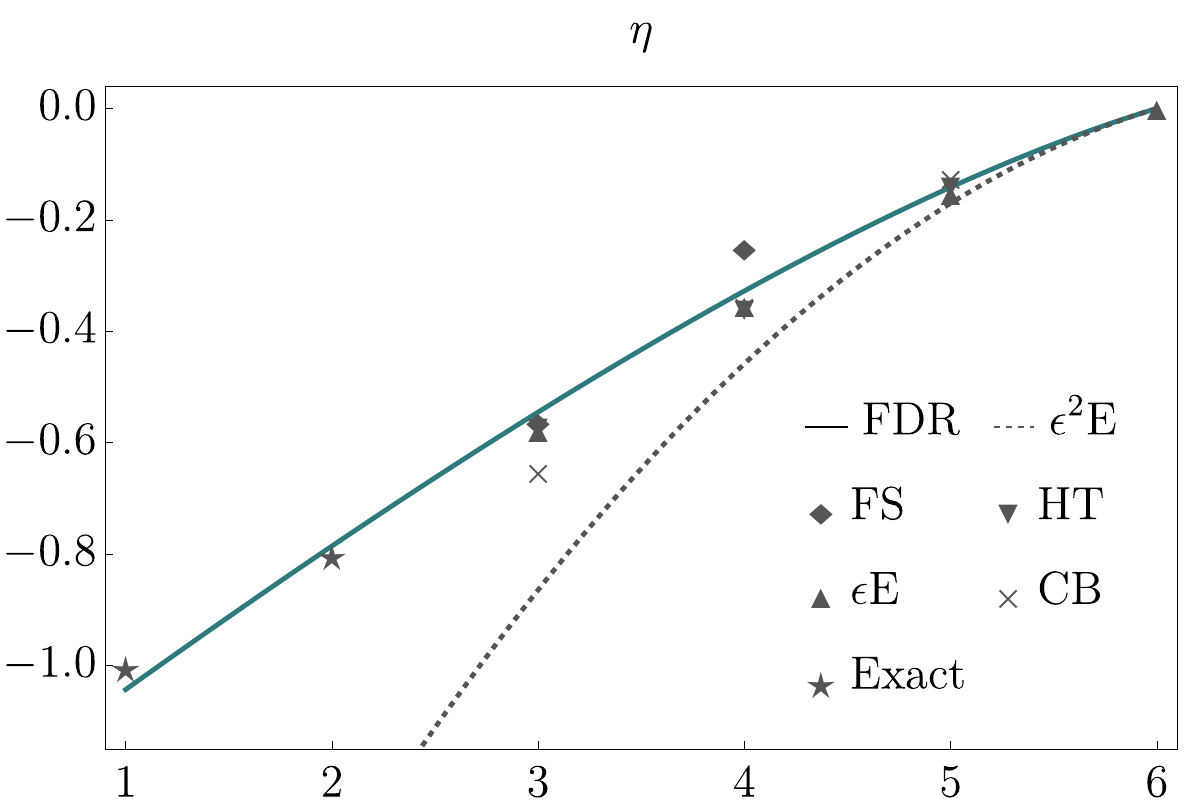}
\caption{Anomalous dimension $\eta$ as a function of $d$ for the \texttt{Lee-Yang} universality class. The comparison is made with the FS (diamonds) \cite{ArguelloCruz:2025zuq}, the resummed six-loop $\varepsilon$E (triangles) \cite{Gracey:2025rnz}, the CB (crosses) \cite{Hikami:2017hwv}, and the HT (inverse triangles) \cite{Butera:2012tq}.
The gray dashed line shows $\eta=-\tfrac{1}{9}\varepsilon-\tfrac{43}{729}\varepsilon^2$ \cite{Gracey:2025rnz}. Exact values  $\eta^{2d} = -4/5$ and $\eta^{1d} = -1$ are included  as stars.}
\label{LeeYangEta}
\end{figure}

As mentioned above, beyond the $\varepsilon$-expansion we proceed numerically. To check the convergence of the field expansions, we compute $\eta$ in $d= 5,4,3,2$ and $1$ for different values of $N_v$, as shown in Fig.~\ref{EtaConvergence}, until convergence of the fourth decimal digit of $\eta$ is attained. As noted in previous studies, the expansion with respect to a non-vanishing point is much faster for the {\tt Lee-Yang} model than the zero-field expansion. This difference is particularly pronounced in $d\leq3$, where we observe that the value of $\eta$ in the latter case converges slowly and convergence cannot be reached in practice. In contrast, the zero-mass expansion is very fast, allowing us to estimate $\eta$ in $d\leq3$.
FDR is efficient: for $\phi_0 \neq 0$, not only are we able to converge in all physical dimensions under study, but convergence is achieved with a polynomial expansion. For instance, in $d=5$ the anomalous dimension is determined to four digits, $\eta=-0.1414$, with just a few more couplings than those in \eqref{expansionv} and \eqref{expansionz}. For $d = 4$, where $\eta = -0.3276$, and $d = 3$, where $\eta = -0.5451$, the addition of only few more couplings is required.
In $d=2$ we find $\eta=-0.7856$, stable to four digits for $N_v\geq16$, and in $d=1$ we find $\eta=-1.0432$, which converges more slowly and is stable to four digits only for $N_v\geq24$ (we pushed the zero-mass expansion up to $N_v=30$). These values lie $1.8\%$ and $4.3\%$ from the exact values $-\tfrac45$ and $-1$, which gives a direct measure of the accuracy of the FDR-DE2 approximation.
Our results for $\eta$, computed with the zero-mass expansion, are shown in Fig.~\ref{LeeYangEta} over the whole range $1\leq d \leq 6$ and are compared there with values obtained by other methods and with the exact results in $d=2$ and $d=1$. Near $d=6$, we observe that all estimates are consistent with the $\varepsilon$-expansion, but as the dimension is lowered the general situation becomes less satisfactory, with a wider spread appearing between the different approaches.

\renewcommand{\arraystretch}{1.3}
\begin{table}[t]
\centering
\footnotesize
\setlength{\tabcolsep}{2pt}
\begin{tabular}{@{}c|c|cccccc|c@{}}
\hline
$d$ & CE & FDR & NPRG & $\varepsilon$E & CB & FS& HT & Exact\\
\hline
\multirow{4}{*}{5} & $\eta$ & $\bf -0.141$ &  $\bf -0.132$ & $\bf-0.152$ & $-0.124$ & $-0.12$& $-0.138$ & $--$\\
 & $\Delta_{\phi}$ & $1.429$ & $1.434$  & $ 1.424$ & $\bf 1.438$ & $\bf 1.44$& $1.431$ & $--$\\
 & $\nu$ & $\bf 0.700$ & $0.697$ & $0.702$ & $0.695$ & $0.69$& $0.699$ & $--$\\
 & $\sigma$ & $0.400$ & $0.402$ & $0.398$ & $0.404$ & $0.40$& $\bf0.401$ & $--$\\
\hline
\multirow{4}{*}{4} & $\eta$ & $\bf -0.328$ & $\bf -0.324$ & $\bf -0.354$ & $-0.354$ & $-0.252$ & $-0.359$ & $--$\\
 & $\Delta_\phi$ & $0.836$ & $0.838$ & $0.823$ & $\bf0.823$ & $\bf0.874$& 0.820 & $--$\\
 & $\nu$ & $\bf 1.196$ & $1.194$ & $1.215$ & $1.215$ & $1.144$ & $1.219$ & $--$\\
 & $\sigma$ & $0.264$ & $0.265$ & $0.259$ & $0.259$ & $0.280$ & $\bf0.258$ & $--$\\
\hline
\multirow{4}{*}{3} & $\eta$ & $\bf -0.545$ & $\bf -0.562$ & $\bf -0.578$ & $-0.651$ & $-0.572$ & $-0.571$ & $--$\\
 & $\Delta_\phi$ & $0.227$ & $0.219$ & $0.211$ & $\bf 0.174$ & $\bf 0.214$ & $0.214$ & $--$\\
 & $\nu$ & $4.397$ & $4.566$ & $4.739$ & $5.736$ & $4.673$ & $4.662$ & $--$\\
 & $\sigma$ & $0.082$ & $0.079$ & $0.076$ & $0.062$ & $0.077$ & $\bf0.077$ & $--$\\
 \hline
 \multirow{4}{*}{2} & $\eta$ & $\bf -0.786$ & $\bf -0.814$ & $--$ & $--$ & $--$ & $-0.797$ & $-0.8$\\
 & $\Delta_\phi$ & $-0.393$ & $-0.407$ & $--$ & $--$ & $--$ & $-0.399$ & $-0.4$\\
 & $\nu$ & $-2.546$ & $-2.457$ & $--$ & $--$ & $--$ & $-2.508$ & $-2.5$\\
 & $\sigma$ & $-0.164$ & $-0.169$ & $--$ & $--$ & $--$ & $\bf-0.166$ & $-\tfrac16$\\
 \hline
 \multirow{4}{*}{1} & $\eta$ & $\bf -1.043$ & $\bf -1.07$ & $--$ & $--$ & $--$ & $--$ & $-1.0$\\
 & $\Delta_\phi$ & $-1.022$ & $-1.04$ & $--$ & $--$ & $--$ & $--$ & $-1.0$\\
 & $\nu$ & $-0.979$ & $-0.97$ & $--$ & $--$ & $--$ & $--$ & $-1.0$\\
 & $\sigma$ & $-0.505$ & $-0.51$ & $--$ & $--$ & $--$ & $--$ & $-0.5$\\
\hline
\end{tabular}

\caption{State-of-the-art and exact critical exponents of the \texttt{Lee-Yang} universality class compared with the present results. The NPRG results are the DE2 taken from \cite{Kockler:2026euo}, which also reports DE4 values; we compare with the second-order results because FDR is employed here at the same DE2 level, so that the comparison isolates the effect of the regularization scheme rather than that of the truncation of the derivative expansion. The values determined directly by each method are shown in bold, while the remaining values are obtained using the scaling relations. For compactness, we round the values to the third decimal digit and omit the errors, which can be found in the respective references where available. The six-loop $\varepsilon$E values are from \cite{Gracey:2025rnz}. The CB values are taken from \cite{Hikami:2017hwv}, the FS values from \cite{ArguelloCruz:2025zuq, Qian:2026fbx}, and the HT results from \cite{Butera:2012tq}.}
\label{tbl:comparissons}
\end{table}
\renewcommand{\arraystretch}{1}

In Table~\ref{tbl:comparissons} we report the values of the leading \texttt{Lee-Yang} critical exponents in integer dimensions together with estimates obtained from other methods. For $\eta$, we report the values obtained from the FP, and the remaining exponents are computed using the scaling relations \eqref{scalingDeltasigma} and \eqref{eqn:scaling-relation}. While a significant spread between the different approaches remains, we are pleased to see a quite satisfactory general agreement that reflects a reasonable understanding of the critical \texttt{Lee-Yang} properties between three and six dimensions.
Several features of the results obtained by other methods, shown in Table~\ref{tbl:comparissons}, are worth mentioning. First, in the case of the CB, two different results have been obtained \cite{Hikami:2017hwv}. At $d = 5$ and $d = 4$, the two results for $\Delta_{\phi}$ are similar, whereas at $d = 3$ they are not; here we cite only one of them, and determining which one is correct requires further study. Second, the FS result at $d = 4$ appears to stand out clearly from the others, but it is probably not fully converged. In Ref.~\cite{ArguelloCruz:2025zuq}, it is shown how $\Delta_\phi$ is converging with the simulation size at $d = 4$, and here we report the last value given there. Third, we take the NPRG--DE2 results in \cite{Kockler:2026euo} to compare the DE2 results, but in this article the derivative expansion was pushed to the next order of the DE with results for $\eta$ that are compatible with the $\varepsilon$E in the range $3\leq d\leq 6$. The difference between FDR-DE2 and NPRG-DE2 is larger for lower dimensions and is of the size expected from the DE truncation error. Fourth, values obtained in \cite{li2011functional} using the BMW approximation of the NPRG are omitted for simplicity, but they agree well with the NPRG-DE2 results of \cite{An:2016lni}.

\begin{figure}[!h]
\centering
\includegraphics[width=1\columnwidth]{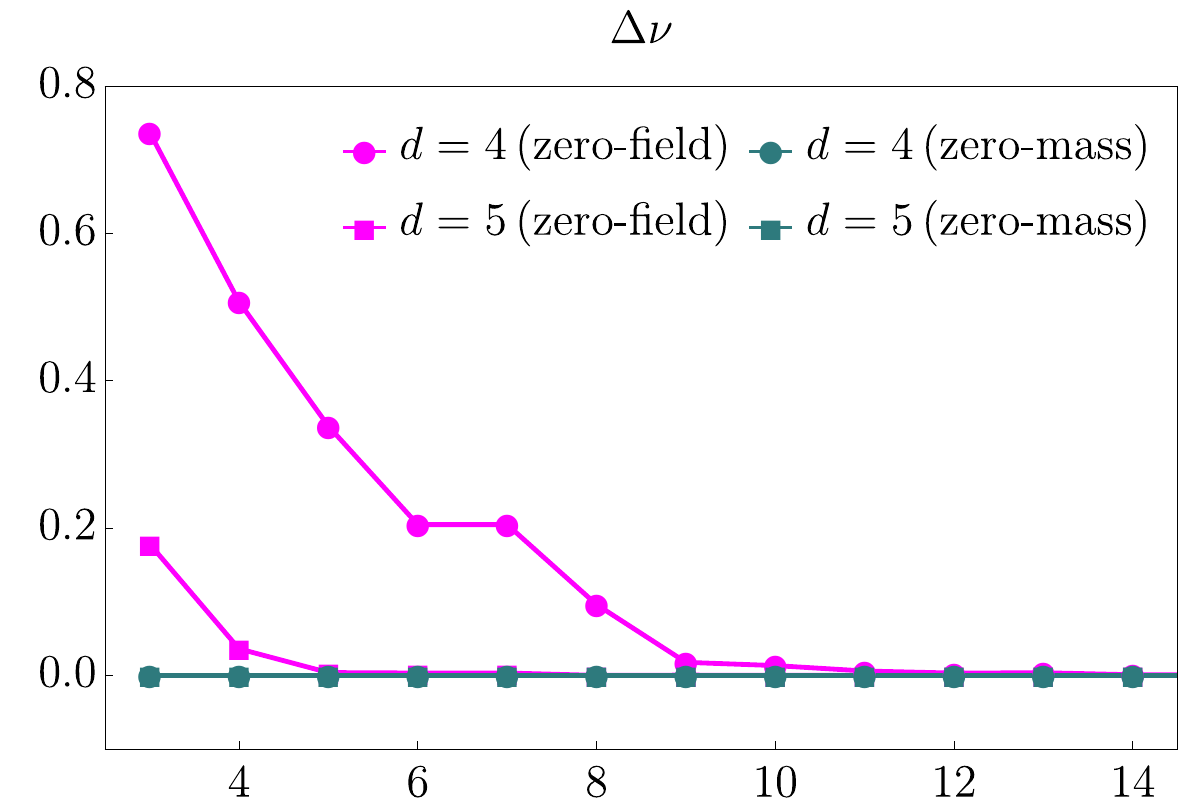}
\caption{$\Delta \nu$ as a function of the truncation order $N_v$ in $d = 4$ and $d = 5$, for both the zero-field expansion and the zero-mass expansion obtained within the FDR-DE2 approximation.  For the zero-mass expansion $\Delta\nu$ vanishes identically.}
\label{fig:deltanus}
\end{figure}
Since the exponent $\nu$ can be computed both from the stability matrix and from the scaling relation \eqref{eqn:scaling-relation}, it is common to define the scaling violation
\begin{equation}
\Delta\nu = \left|\frac{2 \theta_2}{d-2+\eta} +1\right|\,,
\end{equation}
as a measure of the quality of the truncation \cite{Zambelli:2016cbw,An:2016lni}. To test the range of validity of our approximations, we study $\Delta\nu$, following \cite{Boettcher:2015pja,An:2016lni,Zambelli:2016cbw}. Specifically, we compute $\Delta\nu$ for different values of $N_v$ and $d$, as shown in Fig.~\ref{fig:deltanus}. For the expansion around zero-field, we find that $\Delta\nu\rightarrow0$ for sufficiently large values of $N_v$. For the zero-mass expansion the scaling relation is satisfied exactly, $\Delta\nu=0$, at every truncation order and in every dimension, as explained for the NPRG recently in \cite{Kockler:2026euo}.
The scaling violation has been discussed in connection with whether or not the dependence of $\eta$ on the coupling constants is included in the computation of the stability matrix. In \cite{Boettcher:2015pja}, this issue was observed for the decoupled $O(N)+O(M)$ FP, where a field expansion around the minimum of the NPRG flow equations was considered at the LPA' level. In that case, Aharony’s scaling relation is violated when the dependence of $\eta$ on the coupling constants is taken into account in computing $M$, whereas it is satisfied when the FP value of $\eta$ is inserted. A violation of the scaling relation has also been reported in \cite{An:2016lni,Zambelli:2016cbw} when the field expansion is used in the NPRG at DE2 to study the \texttt{Lee-Yang} class. For the expansion around zero-field, the scaling relation is violated regardless of whether the $\eta$ dependence is included in the computation of the stability matrix, but the violation tends to vanish at large orders of the truncation \cite{Zambelli:2016cbw}. As stated above, the zero-mass expansion verifies the scaling relation for all $N_v$ in the NPRG \cite{Kockler:2026euo}.
Using FDR, we obtain results for $\Delta\nu$ that are compatible with the current picture of NPRG at DE2. The violation of the scaling relation for the field expansion around zero-field vanishes for large values of $N_v$. If the $\eta$ dependence is neglected, $\Delta\nu$ does not vanish for the expansion around zero-field. We stress that only when a running $\eta$ is taken into account in the computation of the stability matrix do we obtain the correct order $\varepsilon$ universal result. Note that this effect can only be observed in a theory such as the \texttt{Lee-Yang} model, where $\eta$ is of order $\varepsilon$ in $d=6-\varepsilon$.
A fundamental advantage of FDR compared with NPRG is that the flow equations are simpler, allowing us to push the zero-mass expansion to higher orders than in \cite{Kockler:2026euo}.

\begin{figure}[t!]
\centering
\includegraphics[width=0.98\columnwidth]{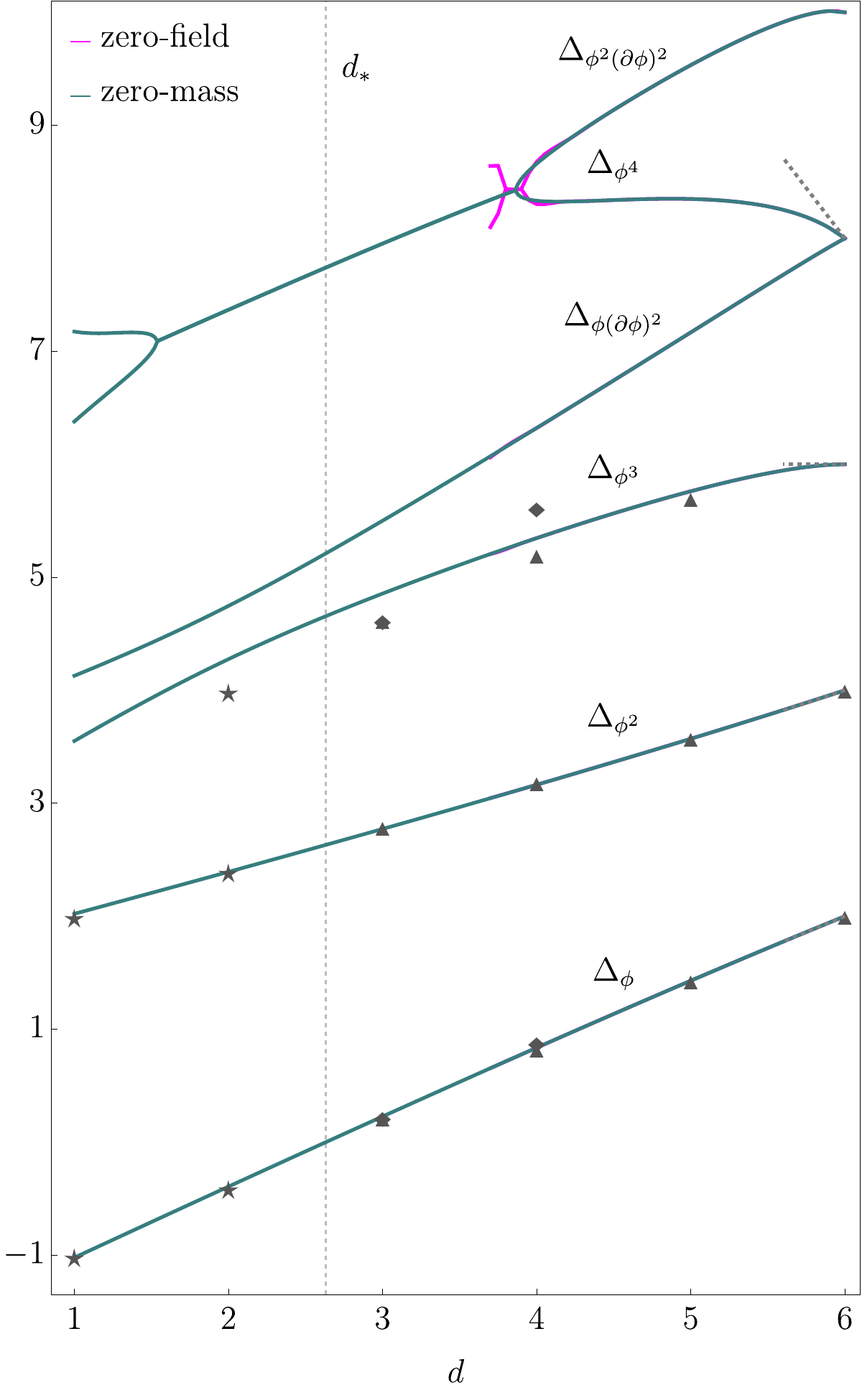}
\caption{The scaling dimensions of spin-$0$ operators with $\Delta<10$ in the \texttt{Lee-Yang} model between the upper critical dimension $d_c=6$ and $d=1$, computed within the FDR-DE2 approximation. Dashed lines represent the leading-order $\varepsilon$-expansion result. For comparison, we also show the values from the $\varepsilon$E to six loops (triangles) \cite{Gracey:2025rnz} and those from the FS (diamonds) \cite{ArguelloCruz:2025zuq}. Exact values are included as stars~\cite{Fisher:1978pf,Cardy:1985yy}.}
\label{LeeYangSpectrum}
\end{figure}

We study the RG spectrum as a function of $d$ for both field expansions. In Fig.~\ref{LeeYangSpectrum} we compare the scaling dimensions \eqref{eqn:scaling-dimension} of the operators appearing in our truncation. The eigenvalue $\theta_2$ vanishes at $d_*\simeq2.63$ and is positive below; this is the dimension marked in Fig.~\ref{LeeYangSpectrum}. To identify each operator, we compare it with the corresponding canonical dimension in $d_c=6$. As can be seen from Fig.~\ref{LeeYangSpectrum}, the scaling dimensions of redundant (inessential) operators \cite{wegner1974some,Baldazzi:2021ydj}, such as $\phi(\partial\phi)^2$, affect our estimates of the spectrum. In the FDR-DE2 approximation studied here, these inessential operators spuriously affect the eigenvalues of the stability matrix. This occurs for either truncation: we observe level repulsion and level-crossing that affects the values of the essential eigenvalues. It is important to stress that we expect the spectrum to consist of real scaling dimensions, as confirmed in $d = 3$ by the CB \cite{Gliozzi:2014jsa} and the FS \cite{ArguelloCruz:2025zuq}. We expect that a reformulation of FDR using the essential RG ideas of \cite{Baldazzi:2021ydj} will solve the problem by eliminating inessential operators from the RG spectrum.
Finally, from Fig.~\ref{LeeYangSpectrum} it is evident that our estimates for $\omega$ are trustworthy in $d=5$, where we find $\omega = 0.7614$ and consequently $\Delta_{\phi^3} = 5.761$, to be compared with  $\Delta_{\phi^3} = 5.697$ from the six-loop $\varepsilon$E of \cite{Gracey:2025rnz}  the FS result $\Delta_{\phi^3}=5.75(2)$ \cite{Qian:2026fbx}. This value is converged: $\omega$ does not change in its fifth digit between $N_v=22$ and $N_v=30$, and the same holds in $d=4$ and $d=3$. The estimates in $d = 4$ and $d = 3$ within the zero-mass expansion are probably shifted by mixing with redundant operators. In any case, we report these estimates: $\Delta_{\phi^3} = 5.346$ in $d=4$ and $\Delta_{\phi^3} = 4.856$ in $d=3$, to be compared with $\Delta_{\phi^3}=5.192$ from $\varepsilon$E in four dimensions \cite{Gracey:2025rnz} and the FS value $\Delta_{\phi^3}=4.613$ in three dimensions \cite{ArguelloCruz:2025zuq}.
Recently, $\omega$ has also been computed with the NPRG at DE4 \cite{Kockler:2026euo}: $\omega=0.701(8)$, $1.20(4)$ and $1.59(3)$ in $d=5,4,3$, in agreement with the six-loop estimates  and, in $d=3$, with the FS. Our FDR-DE2 values lie above the DE4 ones. Since Ref.~\cite{Kockler:2026euo} finds that the corrections from DE2 to DE4 moves $\omega$ towards the six-loop values, the difference may be mainly to the truncation of the derivative expansion. 
For lower dimensions, our results are $\Delta_{\phi^3}=4.275$ in $d=2$ and $\Delta_{\phi^3}=3.551$ in $d=1$ (at $N_v=30$; in $d=1$ the third digit is still uncertain by about $10^{-3}$). These are to be compared with the exact value $\omega=2$ in both $d=2$ and $d=1$ \cite{Fisher:1978pf,Cardy:1985yy,Kockler:2026euo}, which in $d=2$ corresponds to $\Delta_{\phi^3}=4$, and with the NPRG-DE4 result $\omega=1.90(6)$ in $d=2$ \cite{Kockler:2026euo}. The deviation of our values for $\omega$, $14\%$ in $d=2$ and $28\%$ in $d=1$, is consistent with the trend found above $d=3$.
%

\subsection*{\color{teal}Outlook}\vspace{-0.2cm}

In this work, we applied functional dimensional regularization (FDR) to the \texttt{Lee-Yang} universality class using a second-order derivative expansion (DE2). Employing two polynomial truncations -- one around zero-field and one around zero-mass -- we studied the critical theory continuously across dimensions, from the upper critical dimension ($d_c=6$) down to $d=1$.
We studied the convergence of this approach by determining $\eta$ as a function of the polynomial truncation order and of the dimension. We observed that the expansion around a nonzero-field converges faster than the expansion around zero-field, as has also been observed in implementations of the NPRG~\cite{canet2003optimization,Zambelli:2016cbw,An:2016lni,Kockler:2026euo}. In particular, we pushed truncations up to polynomials of degree thirty, which was enough to observe full convergence in the zero-mass expansion, including $d=2$ and $d=1$. We found that both approximations yield the same critical exponent $\eta$ in $d=5$ and $d=4$, while in $d=3$ only the zero-mass expansion had converged. Since the critical exponents of the {\tt Lee-Yang} universality class are related by scaling relations, determining $\eta$ allowed us to estimate $\nu$ and $\sigma$ down to the one-dimensional theory, where the comparison with the exact results in $d=2$ ($1.8\%$ on $\eta$) and $d=1$ ($4.3\%$) quantifies the accuracy of FDR-DE2.
The critical exponents obtained with this approach are competitive with current state-of-the-art calculations. The effectiveness and simplicity of the method suggest that the FDR framework captures essential aspects of the renormalization group flow. This success not only validates the FDR approach but also demonstrates its applicability to non-unitary theories, establishing its utility across a broad range of physical contexts.

Furthermore, we computed the critical exponent $\nu$ from a linear stability analysis of the RG fixed point. The values of $\nu$ obtained from the expansion around zero-field agree with those obtained from the scaling relations. However, owing to the slow convergence of this expansion, it did not allow for an analysis in $d=3$. For the zero-mass expansion the scaling relation is satisfied exactly. The higher part of the spectrum is still affected by mixing with inessential (redundant) operators; to remove them one should implement the principles of the Essential RG \cite{Baldazzi:2021ydj} in our setting.

As already discussed in the main text, we aim to extend this analysis to the {\tt Tri-Lee-Yang} class and to the other non-unitary minimal models, with the long-term goal to understand the correct two-dimensional limits of the entire family of such models -- whose upper critical dimensions are given by $d_c = \tfrac{4 n+2}{2 n-1}$ -- and to clarify which minimal models $\mathcal{M}_{p,q}$ they correspond to. To this end, one likely has to go beyond the polynomial approximation and tackle the full functional structure of the RG flow \eqref{betavz}.
Another immediate extension of this study is to compute the location of the edge singularity, as has been done in \cite{Connelly:2020gwa,Rennecke:2022ohx,Johnson:2022cqv} using the NPRG at order DE2. In particular, since the flow equations obtained from FDR are simpler, the field expansions employed so far can be pushed further. Finally, another direction for future study is the dynamics of the {\tt Lee-Yang} model, which has recently been studied to three-loop order \cite{adzhemyan2026dynamiccriticalexponentyanglee}.

\subsection*{\color{teal}Acknowledgments}\vspace{-0.2cm}

We thank O. Zanusso and V. Skokov for comments and feedback. 
The authors acknowledge financial and compute support from the CSIC grant I+D-2022-22520220100174UD.

\bibliography{referencesLY}

\end{document}